\documentclass[10pt]{iopart}

\usepackage{graphicx,graphics,enumerate,float,cite,color}

\begin{document}

\title{Characterization of a gate-defined double quantum dot in a Si/SiGe nanomembrane}
\date{\today}
\author{T. J. Knapp,$^{1,2}$ R. T. Mohr,$^{2}$ Yize Stephanie Li,$^{3}$ Brandur Thorgrimsson,$^{1,2}$ Ryan H. Foote,$^{1,2}$ Xian Wu,$^{2}$ Daniel R. Ward,$^{2}$ D. E. Savage,$^{3}$ M. G. Lagally,$^{1,3}$ Mark Friesen,$^{1,2}$ S. N. Coppersmith,$^{1,2}$ M. A. Eriksson$^{1,2}$}
\address{$^{1}$Wisconsin Institute for Quantum Information, University of Wisconsin - Madison, 1150 University Avenue, Madison, WI 53706-1390

$^{2}$Department of Physics, University of Wisconsin - Madison, 1150 University Avenue, Madison, WI 53706-1390

$^{3}$Department of Materials Science and Engineering, University of Wisconsin - Madison, 1415 Engineering Drive, Madison, WI 53706-1390}
\ead{tjknapp@wisc.edu}

\begin{abstract}
We report the fabrication and characterization of a gate-defined double quantum dot formed in a Si/SiGe nanomembrane. In the past, all gate-defined quantum dots in Si/SiGe heterostructures were formed on top of strain-graded virtual substrates.  The strain grading process necessarily introduces misfit dislocations into a heterostructure, and these defects introduce lateral strain inhomogeneities, mosaic tilt, and threading dislocations.  The use of a SiGe nanomembrane as the virtual substrate enables the strain relaxation to be entirely elastic, eliminating the need for misfit dislocations. However, in this approach the formation of the heterostructure is more complicated, involving two separate epitaxial growth procedures separated by a wet-transfer process that results in a buried non-epitaxial interface 625 nm from the quantum dot.  We demonstrate that in spite of this buried interface in close proximity to the device, a double quantum dot can be formed that is controllable enough to enable tuning of the inter-dot tunnel coupling, the identification of spin states, and the measurement of a singlet-to-triplet transition as a function of an applied magnetic field.
\end{abstract}

\submitto{\NT}
\maketitle
\ioptwocol

\section{Introduction}

Quantum dots in group-IV semiconductor heterostructures have the potential to be suitable for scalable quantum computing, and have made important steps in recent years towards that goal~\cite{Loss:1998p120,Zwanenburg:2013p961}. Quantum dot qubits in silicon can be formed in several different ways by harnessing a combination of spin and/or charge states: successful realizations have demonstrated the single-spin qubit~\cite{Kawakami:2014p666,Veldhorst:2014p981,Hao:2014p3860,Veldhorst:2015p410}, the singlet-triplet qubit~\cite{Maune:2012p344,Wu:2014p11938}, the charge qubit~\cite{Wang:2013p046801,Shi:2013p075416,Kim:2015p243}, the exchange-only qubit~\cite{Eng:2015p41}, and the hybrid quantum dot qubit~\cite{Kim:2014p70,Kim:2015preprint}.
While metal-oxide-semiconductor devices can confine electrons at the Si-oxide interface independent of the Si strain state, Si/SiGe heterostructures only confine electrons in the Si quantum well if that well is under tensile strain, a state that is typically achieved by epitaxial growth on relaxed SiGe~\cite{Schaffler:1997p1515}.
Strain grading methods enable the growth of such relaxed SiGe buffer layers, allowing the confinement of electrons in a Si quantum well, and the formation of two-dimensional electron gases with very high mobility~\cite{Ismail:1995p1077,Lu:2009p9418}.

However, the goal of making a large array of uniform Si/SiGe quantum dots is still a major challenge that must be overcome if they are to be used in a scalable quantum computer. A fault tolerant quantum computer may require as many as $10^8$ simultaneously tuned qubits~\cite{Fowler:2012p032324}, yet typical strain-graded heterostructures have qubit-affecting inhomogeneities on the length scale of a single qubit~\cite{Evans:2012p5217}. Three types of wafer inhomogeneities have been studied in detail: variation of lateral strain, mosaic structure, and disorder on Si/SiGe interfaces. Strain inhomogeneities will cause variation in the band gap offset~\cite{Sun:2007p10110}, and strain-graded heterostructures have been shown to include $\pm$0.10\% variations in strain over an area of 260~$\mu$m$^2$~\cite{Paskiewicz:2014p4218}. Mosaic structure (i.e., tilting of crystalline lattice vectors) typically varies enough to impact spin qubits over length scales of one micron~\cite{Evans:2012p5217}. In addition, atomic level disorder on the interface between Si and SiGe layers, including single atomic steps, can greatly suppress the singlet-triplet splitting in quantum dots~\cite{Friesen:2006p202106,Goswami:2007p41}.

All three of these qubit-affecting defects are known to be caused by misfit dislocations, defects that are intentionally introduced as a part of the strain grading process~\cite{Mooney:1996p105}. A heterostructure formed through strain grading processes is depicted in Figure 1a. Regardless of how gradually strain is introduced into a heterostructure, strain grading processes will create a buried network of misfit dislocations~\cite{Mooney:1993p3464}. As a consequence, one of the root causes of heterostructure inhomogeneities cannot be eliminated without adopting a new method of strain relaxation.

A liquid release method of elastic strain relaxation has been proposed as a path towards the production of highly uniform heterostructures, because it provides a path to the formation of relaxed SiGe without the introduction of misfit dislocations~\cite{Roberts:2006p40,Paskiewicz:2011p5814,Paskiewicz:2014p4218,Li:2015p4891}. The process begins by growing a layer of SiGe on silicon-on-insulator (SOI) such that its thickness is below the critical thickness necessary to form misfit dislocations. This single crystal SiGe nanomembrane is then released into liquid solution through subsequent HF, KOH, and H$_2$O dips. Free of any rigid substrate, the nanomembrane elastically relaxes to its natural lattice constant. Finally, the nanomembrane is transferred to a new Si handle wafer where it can undergo further epitaxial growth to form the rest of the heterostructure. Figure 1b depicts a heterostructure created using the liquid release method of strain relaxation. Micro-Raman spectroscopy has shown that there is less lateral strain variation in such transferred nanomembrane heterostructures than in conventionally-strain-graded heterostructures~\cite{Paskiewicz:2014p4218}.
%Any remaining heterostructure defects are likely the result of imperfect transfer procedures\cite{Paskiewicz:2014p4218}, which provides a path for future optimization.
Previous work has shown that electron gases formed in transferred nanomembranes have electron mobility above 40,000~cm$^2$/(V$\cdot$s) at a carrier density of 4$\times10^{11}$~cm$^{-2}$~\cite{Li:2015p4891}, a mobility that in principle is high enough to form gate-defined quantum dots~\cite{Simmons:2009p3234}.  However, the nanomembrane transfer process necessarily introduces a non-epitaxial interface between the Si handle wafer and the transferred SiGe nanomembrane, and this interface has the potential to disrupt the tunability and stability of single-electron devices.

\begin{figure*}[t!]
\centering
\includegraphics[width=\textwidth]{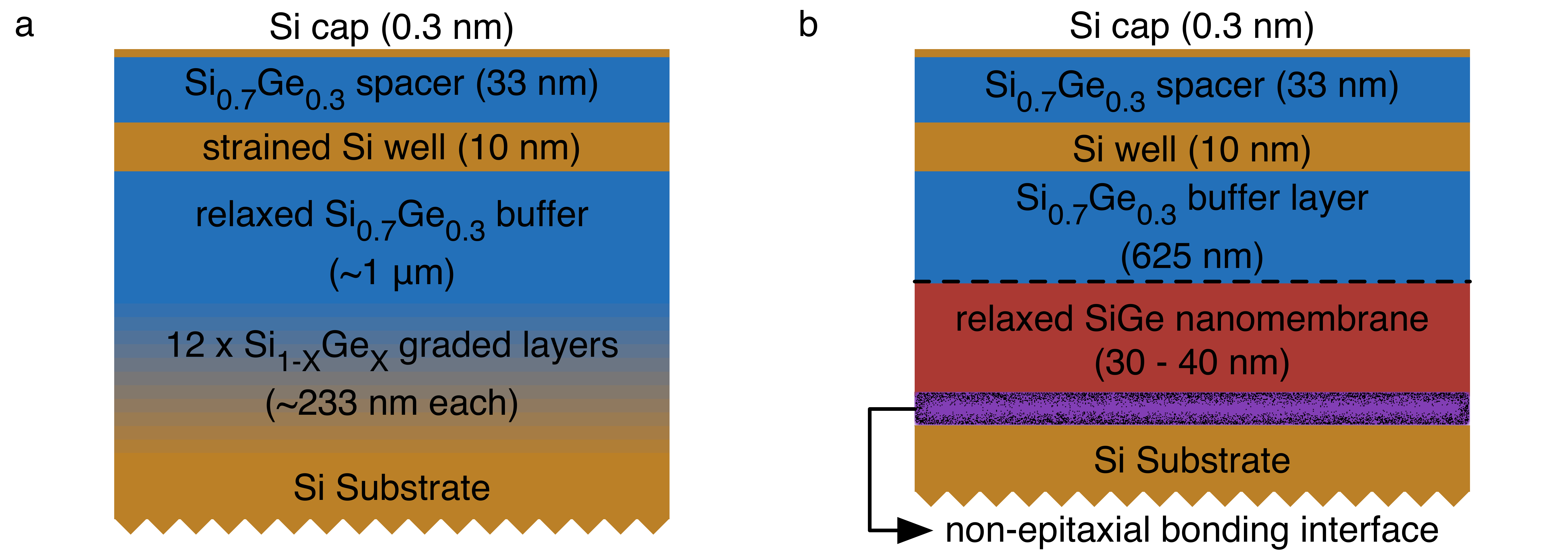}
\caption{Traditional and nanomembrane-based Si/SiGe heterostructures for quantum dots. a) Schematic depiction of a traditional heterostructure, showing the strain-grading method of strain relaxation. Slowly increasing the Ge concentration during growth reduces the density of misfit dislocations and reduces defect bunching. The SiGe buffer, strained Si well, SiGe spacer, and Si cap layers (shown above the graded layers in the diagram) are grown on top of the graded layers. A challenge in working with such strain-graded heterostructures is that the misfit dislocations that are a necessary part of this process contribute to inhomogeneities across an otherwise uniform wafer. b) Schematic depiction of the heterostructure used here, in which the SiGe layer is relaxed elastically using a nanomembrane-based method. The heterostructure below the dashed line is formed by transferring an elastically relaxed SiGe nanomembrane to a new handle wafer, where a non-epitaxial bonding interface is formed. The nanomembrane has no added dislocations, because it was grown below the critical thickness necessary for misfit dislocations, and then was released into liquid solution. After transfer, the SiGe buffer layer, Si quantum well, SiGe spacer, and Si cap were grown by chemical vapor deposition.
}
\end{figure*}

Here we report the first fabrication and characterization of a double quantum dot formed in a Si/SiGe nanomembrane heterostructure. We show that the nanomembrane heterostructure supports the formation and measurement of high-quality, gate-defined quantum dots, including the ability to reach the few electron regime, the ability to coherently tune the inter-dot tunnel coupling of two quantum dots, stability under the application of RF pulses, and sufficient repeatability to enable the performance of magnetospectroscopy of the quantum dot energy levels. These results provide strong evidence that the presence of a non-epitaxial interface between the silicon handle wafer and the wet-transferred nanomembrane does not degrade in an observable way the stability or performance of gate-defined quantum dots.

\section{Methods}
\begin{figure*}[t!]
\centering
\includegraphics[width=\textwidth]{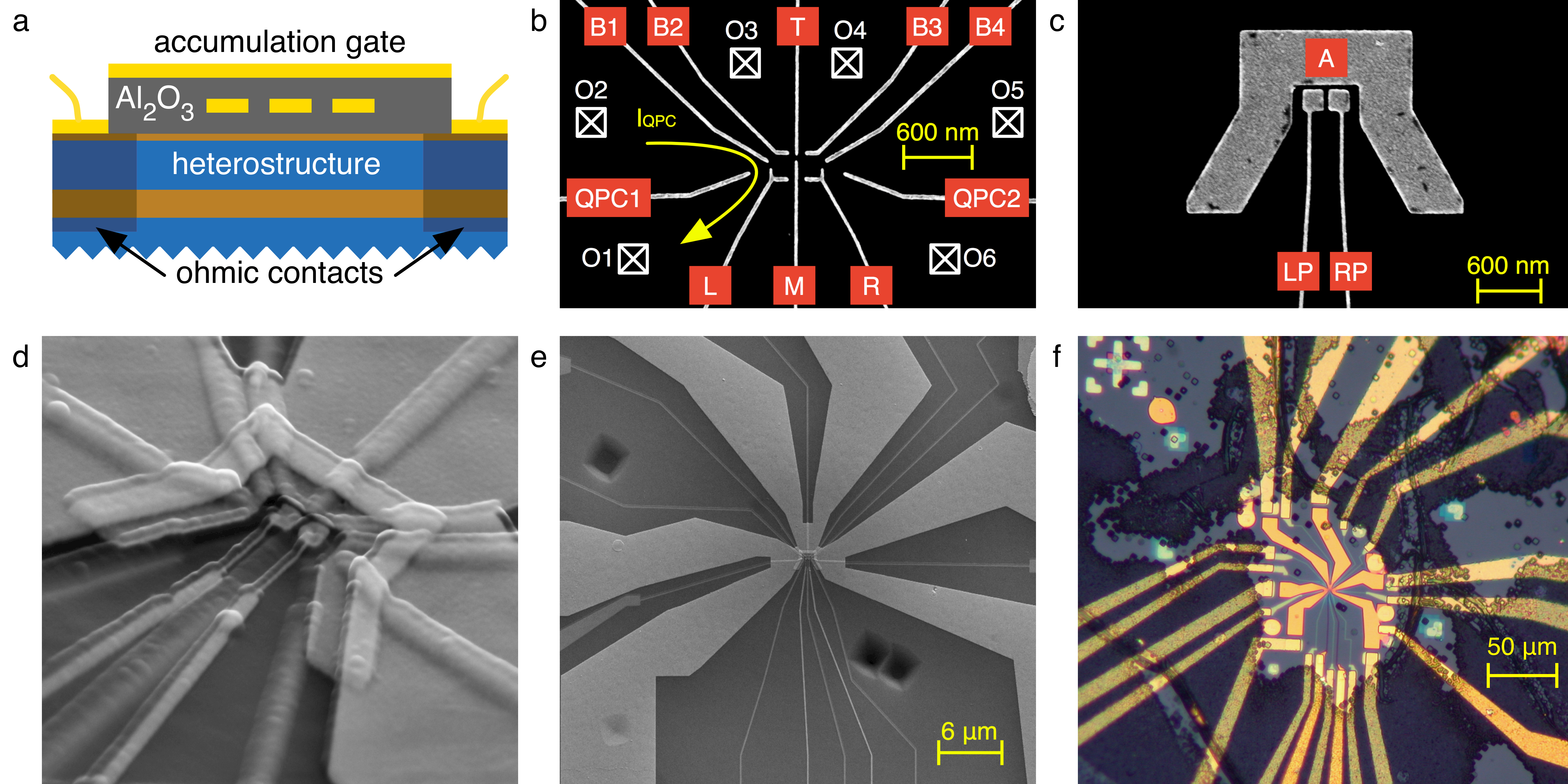}
\caption{Double quantum dot fabricated on a Si/SiGe nanomembrane.
(a) Schematic depiction of a cross-section of the measured device. A 10 nm layer of Al$_2$O$_3$ was added to the heterostructure via atomic layer deposition, and ohmic contacts were created through ion implantation. 
Fine feature depletion gates (Ti/Au) were metallized on top of the oxide; a second layer of Al$_2$O$_3$ was deposited, followed by another layer of Ti/Au gates. 
b) SEM image of a test pattern whose gate design is identical to the lower layer of gates used for the measured device. The lower layer of gates includes two quantum point contacts for charge sensing (QPC1-2), four barrier gates for controlling access to electron reservoirs (B1-4), left and right plunger gates outfitted for RF control of the dot (L and R), and top (T) and middle (M) gates intended to control the inter-dot tunnel rate. The charge sensing experiments reported below were performed with current flowing along the path indicated by the yellow arrow, around QPC1 and through the left quantum point contact.
c) SEM image of a device whose gate design is identical to that of the upper layer of gates of the measured device. The upper layer of gates includes an accumulation gate (A) that controls all of the electron reservoirs, and the left and right paddle (LP and RP, respectively), which control the energy of the left and right dot, respectively. d) Angled SEM image of the measured device, showing the union of two layers of electrostatic gates separated by oxide. No scale bar can be provided because the image was taken at an angle.
e) SEM image of the measured device and the surrounding heterostructure. The gate patterns from panels b and c were placed to avoid visible defects, like the pits seen in this image. Gate A was extended from what is shown in panel c from the dot region all the way to the ohmic contacts. f) Optical image of the measured device, showing rips and folds in the nanomembrane, which appeared during the liquid release and cleaning processes and are discussed in more detail in the text.
}
\end{figure*}

Figure 2a shows a schematic cross section of the Si/SiGe nanomembrane-based double quantum dot studied here, in which all carriers are induced by gates and no dopants are placed in the active area of the device~\cite{Borselli:2011p123118,Ward:2013p213107,Zajac:2015p223507,Borselli:2015p375202}, eliminating a key source of charge noise~\cite{Takeda:2013p123113}. After the second heterostructure growth on the elastically relaxed SiGe nanomembrane, a 10 nm layer of Al$_2$O$_3$ was grown by atomic layer deposition (ALD). To create ohmic contacts, regions of oxide were etched away, and at these locations $^{31}$P donors were implanted with a density of $5\times 10^{15}$~cm$^{-2}$; these regions were subsequently covered with 4~nm Ti and 36~nm Au. The lower layer of gates was patterned using e-beam lithography and metallization with an e-beam evaporator (2~nm Ti/20~nm Au). Figure 2b shows a scanning electron microscope (SEM) image of a test pattern whose gate design is identical to the lower gate layer for the device studied here. This lower layer includes two quantum point contacts for charge sensing (QPC1,2), four barrier gates for controlling access to electron reservoirs (B1-4), left and right plunger gates outfitted for RF control of the dot (L and R), and a top (T) and middle gate (M) intended to control the inter-dot tunnel rate. A second layer of Al$_2$O$_3$ was grown using ALD, this time with a thickness of 80 nm, followed by an upper layer of accumulation gates, which was patterned and metallized with 2~nm Ti and 20~nm Au. Figure 2c shows a test pattern whose gate design is identical to the upper gate layer for the device studied here and includes a left and right paddle for changing dot occupation (PL and PR) and an accumulation gate to regulate access to the ohmic contacts (A). In a final lithography step, this accumulation gate was extended all the way to the ohmic contacts. Figure 2d is a tilt-view SEM image of the completed device.

Figure 2e is a larger-scale, top-down SEM image of the device and the surface of the surrounding heterostructure. Aside from a few pit defects, the heterostructure is quite uniform in the vicinity of the completed double quantum dot, demonstrating that membrane transfer is compatible with smooth and flat heterostructure regions. Figure 2f is an optical image at an even larger scale of the completed device; on this scale numerous tears and gaps are visible in the nanomembrane, and the device was carefully designed and fabricated to fit onto a clean and uniform portion of the nanomembrane.  While this non-uniformity is present in the measured device and made the fabrication more challenging, it does not appear to be an inevitable part of nanomembrane devices; the damage observed in Figure 2f could have been avoided by carefully regulating the sonication power used during the cleaning step immediately prior to the second epitaxial growth. The completed double quantum dot structure was mounted in a dilution refrigerator with a base temperature of $\sim$40~mK.

\section{Results and Discussion}

\begin{figure*}[t!]
\centering
\includegraphics[width=\textwidth]{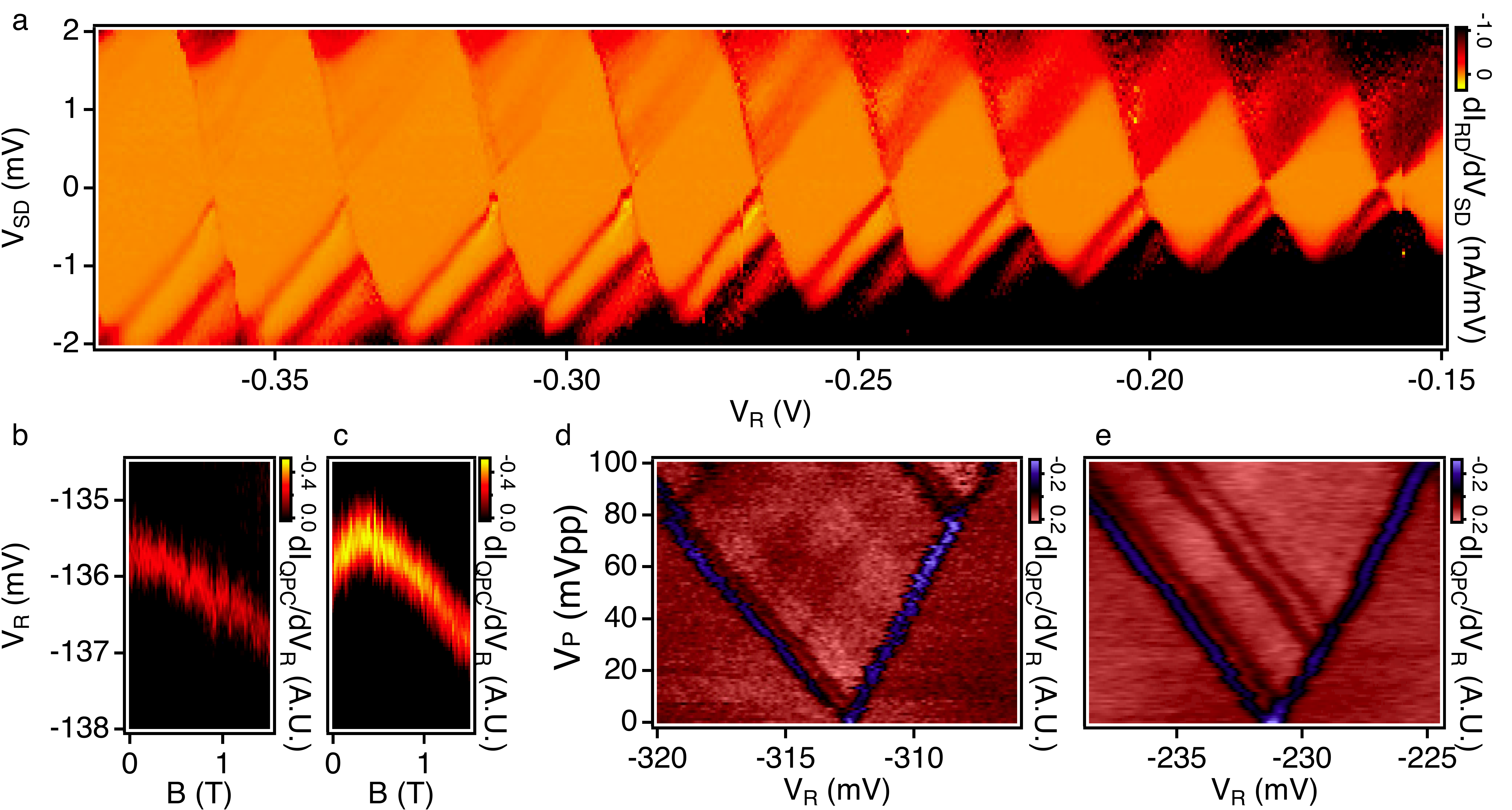}
\caption{Characterization of a single dot formed on the right side of the device, RD. a) Bias spectroscopy in the many electron regime, plotting the differential conductance across RD with current flowing from O3 to O4. Coulomb blockade is manifest with variation of the voltage on gate R ($V_\mathrm{R}$) and the source-drain bias ($V_{\mathrm{SD}}$). Panels b-e present charge-sensing measurements using the left quantum point contact, with current flowing around QPC1 from O1 to O2. Panels b and c show the results of magnetospectroscopy on the 0-to-1 and 1-to-2 electron transitions, respectively. The singlet and triplet two-electron states are degenerate at 0.38 T, which corresponds to a zero-field singlet-triplet splitting of 44~$\mu$eV. Panel c has the same vertical scale range as panel b and is shifted by -75.5~mV. Panels d and e present pulsed-gate spectroscopy on the 0-to-1 electron transition and the 1-to-2 electron transition respectively. Panels d and e are obtained by measuring the differential conductance of the QPC with respect to gate R while pulse trains of peak-to-peak amplitude $V_\mathrm{P}$ are applied to gate R at frequencies 1.0~MHz and 500~kHz, respectively, which enables the characterization of excited states of the dot~\cite{Huebl:2010p1868}.
Panel d shows an excited state 56 $\mu$eV above the 1 electron ground state (likely a valley excitation). Panel e shows an excited state 50 $\mu$eV above the 2 electron state (likely a triplet state where one electron has accessed the valley excited state from panel d).
}
\end{figure*}

\begin{figure*}[t!]
\centering
\includegraphics[width=\textwidth]{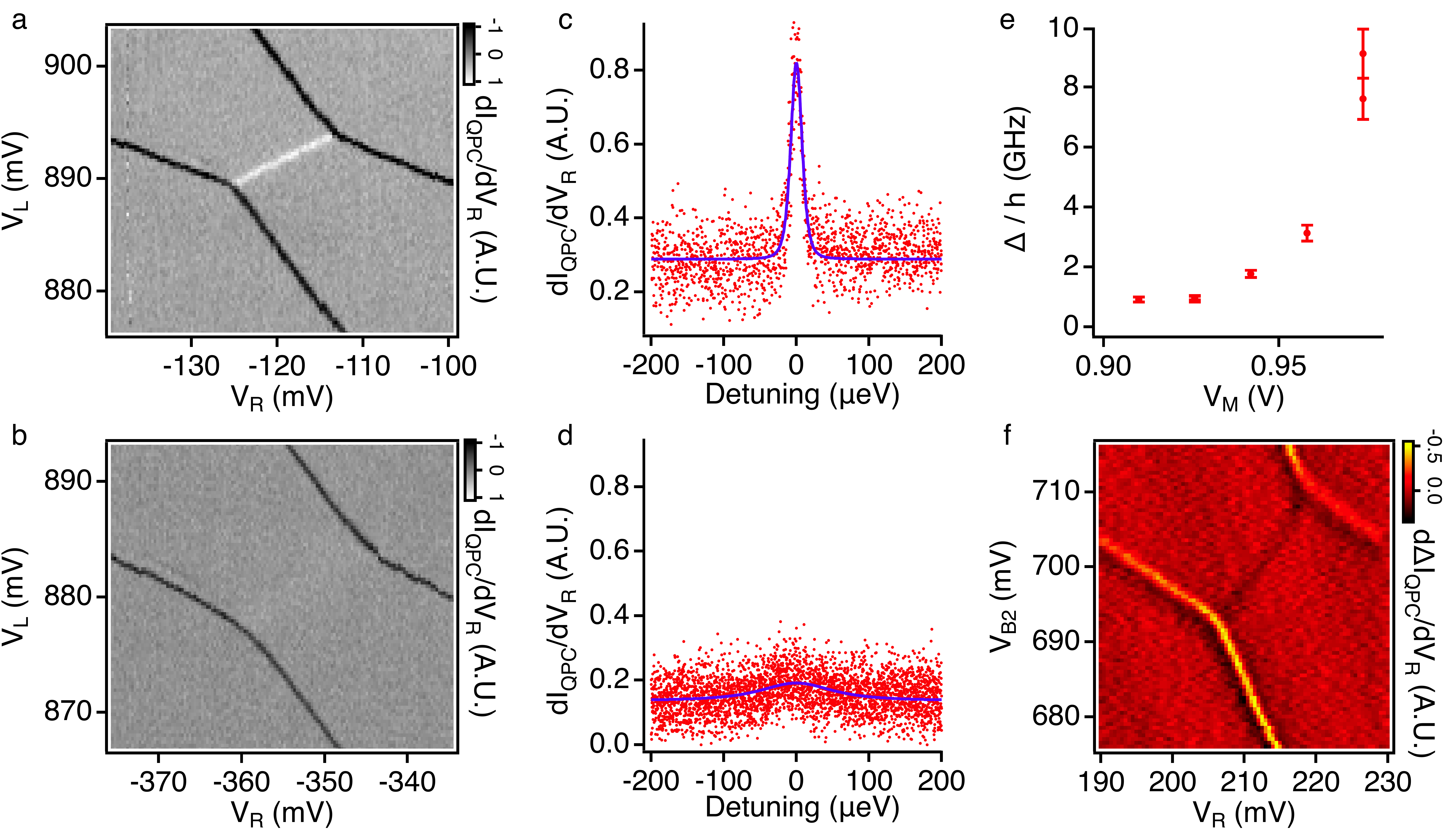}
\caption{Measurements of a double quantum dot formed in a nanomembrane, demonstrating a highly tunable inter-dot tunnel coupling. Charge sensing is performed by modulating the voltage applied to gate R at 213 Hz and performing a lock-in measurement of the real component of the response of the current from O1 to O2. Panels a and b show stability diagrams measured by varying $V_\mathrm{R}$ and $V_\mathrm{L}$.
A negative current response (a dark line) is observed when an electron transitions from either dot to a reservoir, and a positive current response (a white line) occurs at the polarization line at which an electron shifts between the dots.
a) Stability diagram obtained with $\mathrm{V_M}=0.926~V$, where the inter-dot tunnel coupling is small. The polarization line is sharp because the tunnel rate is low and
thus the lifetime broadening is small.
b) Stability diagram obtained with $\mathrm{V_M}=0.974~V$, where the tunnel coupling is high. The polarization line is barely visible because it is strongly lifetime-broadened. 
Panels c) and d): Tunnel couplings are extracted from the stability diagrams in panels a and b by converting measured voltages into detuning using gate lever arms and fitting the transition widths, following Ref.~\cite{Dicarlo:2004p1440}. The gate lever arm $\alpha_\mathrm{R,RD}$ was extracted from the magnetospectroscopy data of Figure 3; other pertinent gate lever arms were then obtained geometrically using the data in panels a and b.
e) Plot of the extracted tunnel coupling $\Delta$ over a range of values for $V_\mathrm{M}$, showing that the tunnel coupling can be tuned over a wide range. The error bars are derived from an unweighted least-squares fit.
f) Charge stability diagram extracted by applying nominally 200 ps square pulses to gate L, which creates a current response $\Delta I$ in the left QPC, thereby demonstrating charge stability under application of high frequency pulses.
}
\end{figure*}

We first present measurements of a single quantum dot formed on the right side of the device (RD), which could be measured both by transport through the quantum dot between ohmic contacts O3 and O4 and by charge-sensing by measuring the current through the left quantum point contact, that is to say around QPC 1 and between ohmic contacts O1 and O2. Figure 3a shows the differential conductance across the quantum dot as a function of the applied source-drain bias voltage and the voltage on gate R.  Numerous Coulomb diamonds are visible with no significant background charge rearrangement, demonstrating the stability of the background charge environment provided by the nanomembrane.

The Coulomb diamonds in Figure 3a increase in size with each expelled electron as gate voltage $V_\mathrm{R}$ is made more negative, moving from right to left in the figure.  Continuing in this direction, the quantum dot was brought into the few-electron regime by tuning the voltages on gates R and RP. Further characterization in the few-electron regime was performed by charge sensing using the left quantum point contact, performing a lock-in measurement of the current between ohmic contacts O1 and O2 while modulating $V_\mathrm{R}$.  Figure 3b shows the results of magnetospectroscopy for a transition where the number of valence electrons changes from zero to one (below we show evidence that there may be an additional closed shell of electrons).
As the in-plane magnetic field $B$ is increased from 0 T to 1.5 T, the transition from zero to one electron occurs at progressively lower values of $V_\mathrm{R}$, consistent with loading a spin-down electron. Figure 3c shows the results of an analogous magnetospectroscopy measurement on the 1-to-2 electron transition; in this case, as the applied magnetic field increases from 0~T to 0.38~T, the transition from one to two electrons occurs at progressively higher values of $V_\mathrm{R}$, consistent with a transition from a single spin-down electron to a two-electron singlet (S) state. In contrast, from $B=0.38$~T to 1.5~T the transition from one to two electrons occurs at progressively lower values of $V_\mathrm{R}$, consistent with the transition from a single spin-down electron to a two-electron triplet T$_-$ state.  The transition of the two-electron ground state from singlet to triplet at $B=0.38$~T corresponds to a zero-field singlet-triplet splitting of 44~$\mu$eV, comparable to values observed in quantum dots grown on conventional strain-graded heterostructures~\cite{Shi:2011p233108,Prance:2012p046808,Maune:2012p344}.

To probe the excited states of the dot, pulsed-gate spectroscopy was performed, with a square wave voltage  applied to gate R enabling loading of excited states~\cite{Elzerman:2004p731,Thalakulam:2011p045307}. The gate lever arms used
to convert the voltage on gate R to the electrostatic energy of the right dot are $\alpha_\mathrm{R,RD}=78$~$\mu$eV/mV for the 0-to-1 electron transition and $\alpha_\mathrm{R,RD}=45$~$\mu$eV/mV for the 1-to-2 electron transition and were extracted from the slope of the transition lines in Figures 3b,c. Figure 3d shows data for the 0-to-1 transition; in this panel, a series of 420~ns square pulses with amplitude $V_\mathrm{P}$ was applied to gate R with a repetition rate of 1 MHz. The single-electron excited state with the lowest energy is 56~$\mu$eV above the ground state, an energy that is consistent with a predominately valley-like excitation in Si/SiGe quantum devices~\cite{Goswami:2007p41,Borselli:2011p123118,Shi:2014p3020,Mi:2015p035304}. In the upper left corner of Figure 3d a line extends towards the lower left and intersects with the so-called ``loading line.''  This line would correspond to an excited state of the 0-electron configuration, and thus it suggests the presence of a closed shell of electrons in the quantum dot.  Figure 3e shows pulsed-gate spectroscopy for the 1-to-2 electron transition. For this transition, a series of 800~ns square pulses with amplitude $V_\mathrm{P}$ was applied to gate R with a repetition rate of 500 kHz, which allowed loading of the two-electron excited states. Three two-electron excited states are visible; the two-electron excited state with the lowest energy sits 50~$\mu$eV above the two-electron ground state, consistent with the singlet-triplet splitting observed in Figure 3c.

We now discuss the formation of a double quantum dot, for which gate M was used to control the inter-dot tunnel rate.  Figures 4a,b show stability diagrams for the device, with Figure 4a acquired with $V_\mathrm{M}=0.926$~V and Figure 4b acquired with $V_\mathrm{M}=0.974$~V. The difference in tunnel rate between these two stability diagrams can be observed in two ways: first, while the polarization line in Figure 4a appears as a sharp, white line, the polarization line in Figure 4b is barely visible because the increased inter-dot tunnel rate leads to strong lifetime broadening. Second, whereas the dot-to-reservoir electron transitions in Figure 4a have sharp corners at their junction with the polarization line, the dot-to-reservoir electron transitions in Figure 4b show significant rounding near junctions with the polarization line.

To extract the tunnel coupling from Figure 4a, Figure 4c plots the lock-in signal as a function of the electrostatic energy difference between the left and right dots (the detuning), effectively superposing sections through the polarization line with many different values for the total double-dot energy. Each data point was projected onto the detuning axis using the pertinent gate lever arms; $\alpha_\mathrm{R, RD}$ was established from the magnetospectroscopy data, and the other lever arms were determined geometrically from the slopes of the transition lines in Figure 4a and the value of $\alpha_\mathrm{R, RD}$. Following the approach of Ref.~\cite{Dicarlo:2004p1440}, we express $I_\mathrm{QPC}$, the current through the charge-sensing QPC, as\begin{equation*}I_\mathrm{QPC} = I_0 + \frac{\Delta I_\mathrm{QPC}}{2} \left[1-\frac{\varepsilon}{\Omega}\tanh \left(\frac{\Omega}{2k_BT} \right) \right] + \Gamma \varepsilon,\end{equation*}
where $\varepsilon$ is the detuning, $\Omega=\sqrt{\epsilon^2+4\Delta^2}$ is the energy difference between the two eigenstates, $k_B$ is the Boltzmann constant, $T$ is the electron temperature, $I_0$ is a current offset fit parameter, $\Delta I_\mathrm{QPC}$ is a parameter for fitting the quantum point contact's sensitivity to an inter-dot charge transition, and $\Gamma$ is a parameter characterizing gate-to-QPC crosstalk. The data from Figures 4c,d are fit with $\partial I_\mathrm{QPC}/\partial V_R=$
\begin{equation*}
\alpha_\mathrm{R,\epsilon}\frac{\Delta I_\mathrm{QPC}}{2}\left[\left(\frac{\epsilon^2}{\Omega^2}-1\right)\frac{\tanh\left(\frac{\Omega}{2k_BT}\right)}{\Omega}\right.
\end{equation*}
\begin{equation*}
\hspace{80pt}\left.+\frac{-\epsilon^2/\Omega^2}{2k_BT\cosh^2\left(\frac{\Omega}{2k_BT}\right)}\right]+\alpha_\mathrm{R,\epsilon}\Gamma,
\end{equation*}
where $\alpha_\mathrm{R,\epsilon}=\partial\epsilon/\partial V_\mathrm{R}$ is the lever arm that converts changes in the voltage on gate R to changes in the detuning. The electron temperature T was taken to be 50~mK, consistent with the width of the dot-to-reservoir transitions (assumed to be primarily temperature broadened). The fits to the data in Figures 4c,d yield $\Delta/\mathrm{h}=$ 0.97$\pm$0.08~GHz and $\Delta/\mathrm{h}=$ 9.1$\pm$0.8~GHz, respectively, where $\Delta$ is the inter-dot tunnel coupling. Figure 4e presents the results of a similar analysis for a total of 7 datasets, demonstrating the achievement of a wide range of tunnel couplings for various values of $V_\mathrm{M}$. The extraction of $\Delta$ is difficult for very large tunnel couplings, as is clear from Figure 4d; for this reason, we plot two values of the tunnel coupling for $V_\mathrm{M}=0.974$~V in Figure 4e, which were extracted from two different stability diagrams taken roughly a day and a half apart. Similarly, two values for the tunnel coupling are plotted for $V_\mathrm{M}=0.926$~V, values which were extracted from separate stability diagrams taken roughly a day apart. In comparison to the data points for $V_\mathrm{M}=0.974$~V, the two data points for $V_\mathrm{M}=0.926$~V are in such close agreement that they are difficult to distinguish.

Finally, Figure 4f reports measurement of a double-dot stability diagram in the presence of short voltage pulses, demonstrating the stability of the device in the presence of very high-bandwidth driving. An arbitrary wave generator with a rise time of 40~ps was used to apply a series of 200~ps square pulses to gate L at a repetition rate of 20~MHz, which caused abrupt changes in the electric potential of the dots. When either dot was energetically close to a charge transition, applying such a series of pulses was likely to induce charge transitions in the dot, which caused a change in QPC current $\Delta I$ relative to a similar sequence of null pulses. Figure 4f shows the derivative of $\Delta I$ with respect to $V_\mathrm{R}$, which highlights regions where a charge transition is induced by the applied sequence of square pulses. The nanomembrane heterostructure of this device provides a stable electrostatic environment for the quantum dots, even when the double quantum dot system is driven by high-bandwidth voltage pulses.

\section{Conclusions}
The results discussed above demonstrate a new path towards the confinement of electrons in Si/SiGe gate-defined quantum dots: we have reported characterization of a double quantum dot formed in a heterostructure created using the liquid release method of strain relaxation. This method of strain relaxation is a powerful tool for the formation of heterostructures with much better uniformity than those created through conventional relaxation methods. The key advantage of this approach is that it does not depend on the insertion of misfit dislocations for strain relaxation, and instead it relies entirely on elastic relaxation of a single-crystal SiGe membrane. The measurements we report here address the stability of a double quantum dot fabricated on a Si/SiGe heterostructure grown epitaxially on such a transferred, relaxed SiGe nanomembrane. We demonstrated that the liquid release method of strain relaxation can produce quantum dots that are stable under a wide range of measurement conditions, and we showed that the inter-dot tunnel coupling was easily tuned over a wide range of values. The buried interface created during the wet transfer of the relaxed nanomembrane is far less controlled than the purely epitaxial heterostructures that in the past were used exclusively for Si/SiGe quantum dot experiments.  The results presented above provide significant evidence that this interface formed 625~nm below the quantum well does not preclude the formation of high-quality and stable double quantum dots.

\section{Acknowledgements}
This work was supported in part by ARO (W911NF-12-0607), NSF (DMR-1206915, PHY-1104660), and the Department of Defense.  
The views and conclusions contained in this document are those of the authors and
should not be interpreted as representing the official policies, either expressly or implied, of the US Government.
Development and maintenance of the growth facilities used for fabricating samples is supported by DOE (DE-FG02-03ER46028).
This research utilized NSF-supported shared facilities at the University of Wisconsin-Madison.

\section{References}
\bibliographystyle{unsrt}
\bibliography{new,main}

\begin{thebibliography}{10}

\bibitem{Loss:1998p120}
D.~Loss and D.~P. DiVincenzo.
\newblock Quantum computation with quantum dots.
\newblock {\em Phys. Rev. A}, 57(1):120--126, Jan 1998.

\bibitem{Zwanenburg:2013p961}
F.~A. Zwanenburg, A.~S. Dzurak, A.~Morello, M.~Y. Simmons, L.~C.~L. Hollenberg,
  G.~Klimeck, S.~Rogge, S.~N. Coppersmith, and M.~A. Eriksson.
\newblock Silicon quantum electronics.
\newblock {\em Rev. Mod. Phys.}, 85:961, 2013.

\bibitem{Kawakami:2014p666}
E.~Kawakami, P.~Scarlino, D.~R. Ward, F.~R. Braakman, D.~E. Savage, M.~G.
  Lagally, M.~Friesen, S.~N. Coppersmith, M.~A. Eriksson, and L.~M.~K.
  Vandersypen.
\newblock Electrical control of a long-lived spin qubit in a {Si/SiGe} quantum
  dot.
\newblock {\em Nature Nanotech.}, 9:666--670, 2014.

\bibitem{Veldhorst:2014p981}
M.~Veldhorst, J.~C.~C. Hwang, C.~H. Yang, A.~W. Leenstra, B.~de~Ronde, J.~P.
  Dehollain, J.~T. Muhonen, F.~E. Hudson, K.~M. Itoh, A.~Morello, and A.~S.
  Dzurak.
\newblock An addressable quantum dot qubit with fault-tolerant
  control-fidelity.
\newblock {\em Nature Nanotech.}, 9(12):981--985, 2014.

\bibitem{Hao:2014p3860}
X.~Hao, R.~Ruskov, M.~Xiao, C.~Tahan, and H.~W. Jiang.
\newblock {Electron spin resonance and spin{\textendash}valley physics in a
  silicon double quantum dot}.
\newblock {\em Nat. Comm.}, 5:3860, 2014.

\bibitem{Veldhorst:2015p410}
M.~Veldhorst, C.~H. Yang, J.~C.~C. Hwang, W.~Huang, J.~P. Dehollain, J.~T.
  Muhonen, S.~Simmons, A.~Laucht, F.~E. Hudson, K.~M. Itoh, A.~Morello, and
  A.~S. Dzurak.
\newblock A two-qubit logic gate in silicon.
\newblock {\em Nature}, 526:410, 2015.

\bibitem{Maune:2012p344}
B.~M. Maune, M.~G. Borselli, B.~Huang, T.~D. Ladd, P.~W. Deelman, K.~S.
  Holabird, A.~A. Kiselev, I.~Alvarado-Rodriguez, R.~S. Ross, A.~E. Schmitz,
  M.~Sokolich, C.~A. Watson, M.~F. Gyure, and A.~T. Hunter.
\newblock Coherent singlet-triplet oscillations in a silicon-based double
  quantum dot.
\newblock {\em Nature}, 481(7381):344--347, 2012.

\bibitem{Wu:2014p11938}
Xian Wu, D.~R. Ward, J.~R. Prance, Dohun Kim, John~King Gamble, R.~T. Mohr,
  Zhan Shi, D.~E. Savage, M.~G. Lagally, Mark Friesen, S.~N. Coppersmith, and
  M.~A Eriksson.
\newblock Two-axis control of singlet-triplet qubit with an integrated
  micromagnet.
\newblock {\em PNAS}, 111:11938, 2014.

\bibitem{Wang:2013p046801}
K.~Wang, C.~Payette, Y.~Dovzhenko, P.~W. Deelman, and J.~R. Petta.
\newblock Charge relaxation in a single-electron {Si}/{SiGe} double quantum
  dot.
\newblock {\em Phys. Rev. Lett.}, 111:046801, 2013.

\bibitem{Shi:2013p075416}
Z.~Shi, C.~B. Simmons, D.~R. Ward, J.~R. Prance, T.~S. Koh, J.~K. Gamble,
  X.~Wu, D.~E. Savage, M.~G. Lagally, M.~Friesen, S.~N. Coppersmith, and M.~A.
  Eriksson.
\newblock Coherent quantum oscillations and echo measurements of a {S}i charge
  qubit.
\newblock {\em Phys. Rev. B}, 88(7):075416, 2013.

\bibitem{Kim:2015p243}
Dohun Kim, D.~R. Ward, C.~B. Simmons, John~King Gamble, Robin Blume-Kohout,
  Erik Nielsen, D.~E. Savage, M.~G. Lagally, Mark Friesen, S.~N. Coppersmith,
  and M.~A. Eriksson.
\newblock Microwave-driven coherent operations of a semiconductor quantum dot
  charge qubit.
\newblock {\em Nature Nanotechnol.}, 10:243--247, 2015.

\bibitem{Eng:2015p41}
Kevin Eng, Thaddeus~D. Ladd, Aaron Smith, Matthew~G. Borselli, Andrey~A.
  Kiselev, Bryan~H. Fong, Kevin~S. Holabird, Thomas~M. Hazard, Biqin Huang,
  Peter~W. Deelman, Ivan Milosavljevic, Adele~E. Schmitz, Richard~S. Ross,
  Mark~F. Gyure, and Andrew~T. Hunter.
\newblock Isotopically enhanced triple-quantum-dot qubit.
\newblock {\em Science Advances}, 1(4), 2015.

\bibitem{Kim:2014p70}
Dohun Kim, Zhan Shi, C.~B. Simmons, D.~R. Ward, J.~R. Prance, Teck~Seng Koh,
  John~King Gamble, D.~E. Savage, M.~G. Lagally, Mark Friesen, S.~N.
  Coppersmith, and M.~A. Eriksson.
\newblock Quantum control and process tomography of a semiconductor quantum dot
  hybrid qubit.
\newblock {\em Nature}, 511:70--74, 2014.

\bibitem{Kim:2015preprint}
Dohun Kim, D.~R. Ward, C.~B. Simmons, D.~E. Savage, M.~G. Lagally, Mark
  Friesen, S.~N. Coppersmith, and M.~A. Eriksson.
\newblock High fidelity resonant gating of a silicon based quantum dot hybrid
  qubit.
\newblock {\em preprint arXiv:1502.03156}, 2015.

\bibitem{Schaffler:1997p1515}
F.~Sch\"{a}ffler.
\newblock High-mobility {S}i and {G}e structures.
\newblock {\em Semicond. Sci. Tech.}, 12(12):1515--1549, 1997.

\bibitem{Ismail:1995p1077}
K.~Ismail, M.~Arafa, K.~L. Saenger, J.~O. Chu, and B.~S. Meyerson.
\newblock Extremely high-electron-mobility in {S}i/{S}i{G}e modulation-doped
  heterostructures.
\newblock {\em Appl. Phys. Lett.}, 66(9):1077--1079, 1995.

\bibitem{Lu:2009p9418}
T.~M. Lu, D.~C. Tsui, C.-H. Lee, and C.~W. Liu.
\newblock Observation of two-dimensional electron gas in a {S}i quantum well
  with mobility of 1.6$\times$10$^6$cm$^2$/{V}s.
\newblock {\em Applied Physics Letters}, 94(18):182102, 2009.

\bibitem{Fowler:2012p032324}
Austin~G. Fowler, Matteo Mariantoni, John~M. Martinis, and Andrew~N. Cleland.
\newblock Surface codes: Towards practical large-scale quantum computation.
\newblock {\em Phys. Rev. A}, 86:032324, 2012.

\bibitem{Evans:2012p5217}
P.~G. Evans, D.~E. Savage, J.~R. Prance, C.~B. Simmons, M.~G. Lagally, S.~N.
  Coppersmith, M.~A. Eriksson, and T.~U. Schülli.
\newblock Nanoscale distortions of {S}i quantum wells in {S}i/{S}i{G}e
  quantum-electronic heterostructures.
\newblock {\em Advanced Materials}, 24(38):5217--5221, 2012.

\bibitem{Sun:2007p10110}
Y.~Sun, S.~E. Thompson, and T.~Nishida.
\newblock Physics of strain effects in semiconductors and
  metal-oxide-semiconductor field-effect transistors.
\newblock {\em Journal of Applied Physics}, 101(10), 2007.

\bibitem{Paskiewicz:2014p4218}
D.~M. Paskiewicz, D.~E. Savage, M.~V. Holt, P.~G. Evans, and M.~G. Lagally.
\newblock Nanomembrane-based materials for group {I}{V} semiconductor quantum
  electronics.
\newblock {\em Scientific Reports}, 4:4218, 2014.

\bibitem{Friesen:2006p202106}
M.~Friesen, M.~A. Eriksson, and S.~N. Coppersmith.
\newblock Magnetic field dependence of valley splitting in realistic
  {S}i/{S}i{G}e quantum wells.
\newblock {\em Appl. Phys. Lett.}, 89:202106, 2006.

\bibitem{Goswami:2007p41}
S.~Goswami, K.~A. Slinker, M.~Friesen, L.~M. McGuire, J.~L. Truitt, C.~Tahan,
  L.~J. Klein, J.~O. Chu, P.~M. Mooney, D.~W. van~der Weide, R.~Joynt, S.~N.
  Coppersmith, and M.~A. Eriksson.
\newblock Controllable valley splitting in silicon quantum devices.
\newblock {\em Nat. Phys.}, 3:41--45, 2007.

\bibitem{Mooney:1996p105}
P.~Mooney.
\newblock Strain relaxation and dislocations in {S}i{G}e/{S}i structures.
\newblock {\em Materials Science and Engineering: R: Reports}, pages 105--146,
  1996.

\bibitem{Mooney:1993p3464}
P.~M. Mooney, F.~K. LeGoues, J.~O. Chu, and S.~F. Nelson.
\newblock Strain relaxation and mosaic structure in relaxed {S}i{G}e layers.
\newblock {\em Applied Physics Letters}, 62(26):3464--3466, 1993.

\bibitem{Roberts:2006p40}
M.~M. Roberts, L.~J. Klein, D.~E. Savage, K.~A. Slinker, M.~Friesen, G.~Celler,
  M.~A. Eriksson, and M.~G. Lagally.
\newblock Elastically relaxed free-standing strained-silicon nanomembranes.
\newblock {\em Nat Mater}, 5:388--393, 2006.

\bibitem{Paskiewicz:2011p5814}
Deborah~M. Paskiewicz, Boy Tanto, Donald~E. Savage, and Max~G. Lagally.
\newblock Defect-free single-crystal {S}i{G}e: A new material from nanomembrane
  strain engineering.
\newblock {\em ACS Nano}, 5(7):5814--5822, 2011.

\bibitem{Li:2015p4891}
Y.~S. Li, P.~Sookchoo, X.~Cui, R.~Mohr, D.~E. Savage, R.~H. Foote, R.~B.
  Jacobson, J.~R. Sánchez-Pérez, D.~M. Paskiewicz, S.~Wu, D.~R. Ward, S.~N.
  Coppersmith, M.~A. Eriksson, and M.~G. Lagally.
\newblock Electronic transport properties of epitaxial {S}i/{S}i{G}e
  heterostructures grown on single-crystal {S}i{G}e nanomembranes.
\newblock {\em ACS Nano}, 9(5):4891--4899, 2015.

\bibitem{Simmons:2009p3234}
C.~B. Simmons, M.~Thalakulam, B.~M. Rosemeyer, B.~J.~Van Bael, E.~K. Sackmann,
  D.~E. Savage, M.~G. Lagally, R.~Joynt, M.~Friesen, S.~N. Coppersmith, and
  M.~A. Eriksson.
\newblock Charge sensing and controllable tunnel coupling in a {S}i/{S}i{G}e
  double quantum dot.
\newblock {\em Nano Lett.}, 9:3234--3238, 2009.

\bibitem{Borselli:2011p123118}
M.~G. Borselli, R.~S. Ross, A.~A. Kiselev, E.~T. Croke, K.~S. Holabird, P.~W.
  Deelman, L.~D. Warren, I.~Alvarado-Rodriguez, I.~Milosavljevic, F.~C. Ku,
  W.~S. Wong, A.~E. Schmitz, M.~Sokolich, M.~F. Gyure, and A.~T. Hunter.
\newblock Measurement of valley splitting in high-symmetry {S}i/{S}i{G}e
  quantum dots.
\newblock {\em Appl. Phys. Lett.}, 98(12):123118, 2011.

\bibitem{Ward:2013p213107}
D.~R. Ward, D.~E. Savage, M.~G. Lagally, S.~N. Coppersmith, and M.~A. Eriksson.
\newblock Integration of on-chip field-effect transistor switches with
  dopantless {Si/SiGe} quantum dots for high-throughput testing.
\newblock {\em Applied Physics Letters}, 102:213107, 2013.

\bibitem{Zajac:2015p223507}
D.~M. {Zajac}, T.~M. {Hazard}, X.~{Mi}, K.~{Wang}, and J.~R. {Petta}.
\newblock {A reconfigurable gate architecture for {S}i/{S}i{G}e quantum dots}.
\newblock {\em Applied Physics Letters}, 106:223507, 2015.

\bibitem{Borselli:2015p375202}
M.~G. Borselli, K.~Eng, R.~S. Ross, T.~M. Hazard, K.~S. Holabird, B.~Huang,
  A.~A. Kiselev, P.~W. Deelman, L.~D. Warren, I.~Milosavljevic, A.~E. Schmitz,
  M.~Sokolich, M.~F. Gyure, and A.~T. Hunter.
\newblock {Undoped accumulation-mode Si/SiGe quantum dots}.
\newblock {\em Nanotechnology}, 26:375202, 2015.

\bibitem{Takeda:2013p123113}
K.~Takeda, T.~Obata, Y.~Fukuoka, W.~M. Akhtar, J.~Kamioka, T.~Kodera, S.~Oda,
  and S.~Tarucha.
\newblock {Characterization and suppression of low-frequency noise in Si/SiGe
  quantum point contacts and quantum dots}.
\newblock {\em Applied Physics Letters}, 102:123113, 2013.

\bibitem{Huebl:2010p1868}
Hans Huebl, Christopher~D. Nugroho, Andrea Morello, Christopher~C. Escott,
  Mark~A. Eriksson, Changyi Yang, David~N. Jamieson, Robert~G. Clark, and
  Andrew~S. Dzurak.
\newblock Electron tunnel rates in a donor-silicon single electron transistor
  hybrid.
\newblock {\em Phys Rev B}, 81(23):235318, 2010.

\bibitem{Dicarlo:2004p1440}
L.~{DiCarlo}, H.~J. Lynch, A.~C. Johnson, L.~I. Childress, K.~Crockett, and
  C.~M. Marcus.
\newblock Differential charge sensing and charge delocalization in a tunable
  double quantum dot.
\newblock {\em Phys. Rev. Lett.}, 92(22):226801, 2004.

\bibitem{Shi:2011p233108}
Z.~Shi, C.~B. Simmons, J.~R. Prance, John~King Gamble, Mark Friesen, D.~E.
  Savage, M.~G. Lagally, S.~N. Coppersmith, and M.~A. Eriksson.
\newblock Tunable singlet-triplet splitting in a few-electron {Si/SiGe} quantum
  dot.
\newblock {\em Appl. Phys. Lett.}, 99:233108, 2011.

\bibitem{Prance:2012p046808}
J.~R. Prance, Zhan Shi, C.~B. Simmons, D.~E. Savage, M.~G. Lagally, L.~R.
  Schreiber, L.~M.~K Vandersypen, Mark Friesen, Robert Joynt, S.~N.
  Coppersmith, and M.~A. Eriksson.
\newblock Single-shot measurement of triplet-singlet relaxation in a
  {S}i/{S}i{G}e double quantum dot.
\newblock {\em Phys. Rev. Lett.}, 108:046808, 2012.

\bibitem{Elzerman:2004p731}
J.~M. Elzerman, R.~Hanson, L.~H. Willems~van Beveren, L.~M.~K. Vandersypen, and
  L.~P. Kouwenhoven.
\newblock Excited-state spectroscopy on a nearly closed quantum dot via charge
  detection.
\newblock {\em Appl. Phys. Lett.}, 84:4617--4619, 2004.

\bibitem{Thalakulam:2011p045307}
Madhu Thalakulam, C.~B. Simmons, B.~J. Van~Bael, B.~M. Rosemeyer, D.~E. Savage,
  M.~G. Lagally, Mark Friesen, S.~N. Coppersmith, and M.~A. Eriksson.
\newblock Single-shot measurement and tunnel-rate spectroscopy of a
  {S}i/{S}i{G}e few-electron quantum dot.
\newblock {\em Physical Review B}, 84(4):045307, 2011.

\bibitem{Shi:2014p3020}
Zhan Shi, C.~B. Simmons, Daniel~R. Ward, J.~R. Prance, Xian Wu, Teck~Seng Koh,
  John~King Gamble, D.~E. Savage, M.~G. Lagally, Mark Friesen, S.~N.
  Coppersmith, and M.~A. Eriksson.
\newblock Fast coherent manipulation of three-electron states in a double
  quantum dot.
\newblock {\em Nature Comm.}, 5(3020):3020, 2014.

\bibitem{Mi:2015p035304}
X.~Mi, T.~M. Hazard, C.~Payette, K.~Wang, D.~M. Zajac, J.~V. Cady, and J.~R.
  Petta.
\newblock Magnetotransport studies of mobility limiting mechanisms in undoped
  {S}i/{S}i{G}e heterostructures.
\newblock {\em Phys. Rev. B}, 92:035304, 2015.

\end{thebibliography}
\end{document}